\magnification  1095

\baselineskip  12 pt



\def\ex#1{  \medskip  \noindent  {\bf Example #1 }}\medskip
\def \sect#1{\bigskip \noindent {\bf #1}\medskip}
\def \subsect#1{\medskip \noindent {\it #1} \medskip}

\def \th #1#2{\medskip \noindent{\bf Theorem #1}  \it  #2  \rm \medskip}
\def \cor #1#2{\medskip \noindent{\bf Corollary #1}  \it  #2  \rm \medskip}

\def \lem #1#2{\medskip \noindent{\bf Lemma #1}  \it  #2  \rm \medskip}
\def \defi #1#2{\medskip \noindent{\bf Definition #1} #2}
\def \pf {\noindent {\it Proof.\quad}}

\def \noi {\noindent}

\def \la {\lambda}
\def \a {\alpha}

\def \E {{\bf E}}
\def \P {{\bf P}}

\def \rp {{\bf R}_+}

\def \sqr#1#2{{\vcenter{\vbox{\hrule height.#2pt\hbox{\vrule width.#2pt height
#1pt \kern#1pt\vrule width.#2pt}\hrule height.#2pt}}}}

\def\square {\sqr65 }
 \def \qed{ \hfill $\square$}

\centerline{\bf Mutual Fund Theorems when Minimizing the Probability of Lifetime Ruin} \bigskip

\centerline{Version:  19 March 2008}

\bigskip

\noindent Erhan Bayraktar, erhan@umich.edu \hfill \break
\noindent Virginia R. Young, vryoung@umich.edu \hfill \break
\indent Department of Mathematics \hfill \break
\indent University of Michigan \hfill \break
\indent 530 Church Street \hfill \break
\indent Ann Arbor, Michigan, 48109 

\bigskip

\noindent{\bf Abstract:} We show that the mutual fund theorems of Merton (1971) extend to the problem of optimal investment to   minimize the probability of lifetime ruin.  We obtain two such theorems by considering a financial market both with and without a riskless asset for random consumption.  The striking result is that we obtain two-fund theorems despite the additional source of randomness from consumption.  

\medskip

\noi {\bf JEL Classification:} Primary G11; Secondary C61.

\medskip

\noi {\bf MSC 2000 Classification:} Primary 93E20; Secondary 91B28.  

\medskip

\noindent{\bf Keywords:} probability of ruin, mutual fund, separation theorem, optimal investment, Hamilton-Jacobi-Bellman equation.

\sect{1. Introduction}

The contribution of this paper is two-fold.  First, we show that mutual fund theorems hold when minimizing the probability of lifetime ruin (that is, wealth reaching zero before death), as Merton (1971) does when maximizing the utility of consumption.  Bayraktar and Young (2007a) determine when the investment strategies are identical under the two problems of maximizing utility of consumption or minimizing the probability of lifetime ruin.  They show that a necessary condition is that the utility exhibit hyperbolic risk aversion, a commonly used utility function in mathematical finance.  Therefore, the present paper complements the work of Bayraktar and Young (2007a) in further relating the areas of utility maximization and ruin probability minimization by showing that mutual fund theorems hold for the latter problem.  See Bayraktar and Young (2007b) for motivation and additional references concerning minimizing the probability of lifetime ruin as a criterion for optimization.

Second, we show that when consumption is random, then the optimal investment strategy can be expressed in terms of investing in two risky mutual funds--that is, a two-fund theorem holds with both funds risky.  Then, once the problem is reduced to minimizing the probability of lifetime ruin in the presence of two risky assets, one can refer to Bayraktar and Young (2007c) who provide the solution to this problem.  Stochastic income in the utility setting was considered by Duffie et al.\ (1997), but they do not obtain a mutual fund result.

We do not obtain these mutual fund theorems in the most general setting; see Khanna and Kulldorff (1999) for more general mutual fund theorems when maximizing expected utility.  For clarity of presentation, we apply the financial market of Bj\"ork (2004, Section 19.7), although we fully expect that mutual fund theorems hold when minimizing the probability of lifetime ruin in the financial markets of Khanna and Kulldorff (1999).

The remainder of the paper is organized as follows:  In Section 2, we describe the financial market in which the decision maker invests, namely, $n$ risky assets either with or without a riskless asset; we define the probability of lifetime ruin; and we define what we mean by a {\it relative portfolio vector} and the corresponding mutual fund.

In Section 3, we consider the case for which the consumption rate follows a diffusion.  In Section 3.1, we assume that individual cannot invest in a riskless asset and show that a two-fund theorem holds with both funds risky.  In Section 3.2, we include a riskless asset in the financial market; again, we show that a two-fund theorem holds, in which the two mutual funds include the riskless asset in addition to the $n$ risky assets.

It is surprising to us that the random consumption did not result in {\it three}-fund theorems due to the extra source of uncertainty.  Indeed, Merton (1973, Theorem 2) shows that when the drifts and volatilities of the risky assets are functions of a random short rate, then three funds are needed to account for the additional randomness of the short rate.  Under Merton's (1973, Section 7) financial market, we would also obtain a three-fund result in our setting. 

In the special case in which the rate of consumption is deterministic, one of the mutual funds in Section 3.2 can be taken to be the riskless asset, and the resulting mutual funds (in both Sections 3.1 and 3.2) are {\it identical} to those obtained when maximizing expected utility of consumption (Merton, 1971).  Moreover, if the financial market were as in Merton (1973, Section 7), then we would obtain the identical three funds as in Theorem 2 of that paper.

\sect{2. Financial Market and Probability of Lifetime Ruin}

In this section, we present the financial market, namely $n$ risky assets either with or without a riskless asset.  We assume that the individual invests in a given market to minimize the probability that her wealth reaches zero before she dies, the so-called {\it probability of lifetime ruin}.  In Section 3, we show that the optimal investment strategy can be represented as a choice of how much to invest in two mutual funds, as in Merton (1971).

We follow the financial market model in Bj\"ork (2004, Section 19.7).  The individual invests in $n$ risky assets with price vector $S = (S_1, S_2, \dots, S_n)^T$, in which $S_i$ is the price of the $i$th risky asset for $i = 1, 2, \dots, n$, and $T$ denotes the transpose operator.  Assume that $S$ follows the process given by
$$
dS(t) = D(S(t)) \left(\mu(t) dt + \sigma(t) dB(t) \right), \eqno(2.1)
$$
in which $B = (B_1, B_2, \dots, B_k)^T$ is a $k$-dimensional standard Brownian motion with respect to a filtered probability space $(\Omega, {\cal F}, \{{\cal F}(t) \}_{t \ge 0}, \P)$ satisfying the usual assumptions.   The vector $\mu(t) = (\mu_1(t), \mu_2(t), \dots, \mu_n(t))^T$ is an $n$-vector of drifts, and $\sigma(t)$ is an $n \times k$ volatility matrix, deterministic with respect to time.  Finally, $D(S)$ is the diagonal matrix
$$
D(S) = {\rm diag}(S_1, S_2, \dots, S_n).
$$
Alternatively, one can write (2.1) as
$$
dS_i(t) = S_i(t) \left(\mu_i(t) dt + \sigma_i(t) dB(t) \right), \quad i = 1, 2, \dots, n,
$$
in which $\sigma_i(t)$ is the $i$th row of the matrix $\sigma(t)$.  Assume that $\sigma(t)$ is of rank $n$ so that the $n \times n$ variance-covariance matrix $\Sigma(t) := \sigma(t) \sigma(t)^T$ is positive definite and invertible.

We assume that the individual consumes at a random continuous rate $c(t)$ at time $t$; this rate is net of (possibly random) income.  If  the consumption rate is also given net of inflation, then subtract the inflation rate from $\mu_i(t)$ (and in Section 3.2, from the riskless rate $r$, too).  Specifically, we assume that $c$ follows the diffusion given by
$$
dc(t) = c(t)(a(t) \, dt + b(t) \, dB^c(t)), \eqno(2.2)
$$
in which $a(t)$ and $b(t)$ are deterministic functions of time, and $B^c$ is a standard Brownian motion with respect to the filtered probability space $(\Omega, {\cal F}, \{{\cal F}(t)\}_{t \ge 0}, \P)$.  We assume that $B^c$ and $B$ are correlated Brownian motions with correlation coefficient between $B^c$ and $B_i$ denoted by $\rho_i$ for $i  = 1, 2, \dots, k$.  Write $\rho$ for the $k$-vector $(\rho_1, \rho_2, \dots, \rho_k)^T$.

Let $W(t)$ be the wealth at time $t$ of the individual, and let $\pi(t) = (\pi_1(t), \pi_2(t), \dots, \pi_n(t))$ be the {\it amounts} of wealth that the decision maker invests in the risky assets $1, 2, \dots, n$, respectively, at that time.   In Section 3.1, we assume that there is {\it no} riskless asset in which the individual can invest, so we impose the condition that 
$$
e^T \pi(t) = W(t), \eqno(2.3)
$$
for all $t \ge 0$, in which $e = (1, 1, \dots, 1)^T$ is the $n$-vector of 1's.  In Section 3.2, we assume that there {\it is} a riskless asset with constant return of $r \ge 0$ in which the individual can invest, so we do not impose the condition in (2.3).  Thus, when there is a riskless asset (Section 3.2), wealth follows the process
$$
dW(t) = \left( r W(t) + \pi(t)^T (\mu(t) - r e) - c(t) \right) dt +  \pi(t)^T \sigma(t) \, dB(t), \quad W_0 = w > 0. \eqno(2.4)
$$
When there is no riskless asset, as in Section 3.1, then impose the constraint $e^T \pi(t) = W(t)$, or formally set $r = 0$ in (2.4).

Define a hitting time $\tau_0$ associated with the wealth process by $\tau_0 = \inf \{  t\ge 0: W(t) \le 0 \}$.  This hitting time is the time of ruin.  Also, define the random time of death of the individual by $\tau_d$.  We represent $\tau_d$ as the first jump time of a non-homogeneous Poisson process $N(t)$ with deterministic hazard rate $\la(t)$; that is, $\P(N(t) = 0) = \exp(-\int_0^t \la(s) ds)$.  Assume that $\tau_d$ is independent of the $\sigma$-algebra generated by the Brownian motion $B$.

By the {\it probability of lifetime ruin}, we mean the probability that wealth reaches 0 before the individual dies, that is, $\tau_0 < \tau_d$.  We minimize the probability of lifetime ruin with respect to the set of admissible investment strategies ${\cal A}$.  A strategy $\pi$ is {\it admissible} if it is adapted to the filtration ${\{ {\cal F}(t) \}}$ and if it satisfies the integrability condition $\int_0^t  \pi(s)^T \pi(s) \, ds < \infty$, almost surely, for all $t \ge 0$.

Because consumption is random, the probability of lifetime ruin has two state variables, namely, the rate of consumption and the wealth at time $t$.  Thus, the probability of lifetime ruin $\psi$ is defined on $\rp^3$ by
$$
\psi(w, c, t) = \inf_{\pi \in {\cal A}} {\bf P}^{w, c, t} \left( \tau_0 < \tau_d \big| \min(\tau_d, \tau_0) > t \right),\eqno(2.5)
$$
in which ${\bf P}^{w, c, t}$ denotes the conditional probability given $W(t) = w > 0$ and $c(t) = c > 0$, and we also explicitly condition on the individual being alive at time $t$ and not having ruined before then.

Before moving on to the mutual fund theorems in Section 3, we show how a vector of numbers that add to 1 can determine a mutual fund. 

\defi{2.1}{Suppose $g(t) = (g_1(t), g_2(t), \dots, g_n(t))^T$ is an $n$-vector of deterministic functions of time such that $e^T g(t) = 1$ for all $t \ge 0$.  Then, we call $g$ a {\it relative portfolio vector} because it naturally determines a mutual fund in the $n$ risky assets.  Specifically, the mutual fund determined by $g$ is such that the proportion of the fund invested in the $i$th risky asset at time $t$ is given by $g_i(t)$ for $i = 1, 2, \dots, n$; note that this mutual fund requires continual rebalancing in order to maintain the proportions $g_i(t)$.  Then, the price $S^g$ of this mutual fund follows the process
$$dS^g(t) = S^g(t) \left( \mu^g(t) \, dt + \sigma^g(t) \, dB(t) \right), \eqno(2.6)$$
\noi in which $\mu^g(t) := g(t)^T \mu(t)$ and $\sigma^g(t) := g(t)^T \sigma(t)$.}

\vfill \eject

\sect{3. Mutual Fund Theorems}

In Section 3.1, there is no riskless asset in the financial market, while in Section 3.2 there is a riskless asset.  In both cases, the individual's optimal investment strategy is to invest in two mutual funds.  It is surprising to us that the random consumption does not result in {\it three}-fund theorems due to the extra source of uncertainty.  By contrast to the result when consumption is deterministic and there is a riskless asset, in Section 3.2, {\it neither} of the mutual funds can be taken equal to the riskless asset; however, both of these mutual funds contain the riskless asset.  Additionally, to consider the special case for which the consumption rate is deterministic, we set $b$ to be identically zero in (2.2), and we do this to obtain corollaries of our mutual fund theorems.

\subsect{3.1 No Riskless Asset}

When there is no riskless asset, we have the following lemma for the minimum probability of lifetime ruin $\psi$, whose proof follows from the same techniques used by Bayraktar and Young (2007c) in proving Theorem 2.2 of that paper, which considers the problem of minimizing the probability of lifetime ruin in a market with two risky assets and with time-homogeneous parameters.

\lem{3.1}{For the model in this section, the minimum probability of lifetime ruin $\psi$ given in $(2.5)$ is decreasing and convex with respect to $w$, increasing with respect to $c$, and lies in ${\cal C}^{2,2,1}(\rp^3)$.  Additionally, $\psi$ is the unique classical solution of the following Hamilton-Jacobi-Bellman $($HJB$\, )$ equation on $\rp^3:$
$$
\left\{ \eqalign{&  \la(t) \, v = v_t - c \, v_w + a(t) \, c \, v_c + {1 \over 2} \, b^2(t) \, c^2 \, v_{cc} \cr
& \qquad  + \min_{e^T \pi = w} \left[ \pi^T \mu(t) \, v_w + {1 \over 2} \, \pi^T \, \Sigma(t) \, \pi \, v_{ww} + b(t) \, c \, \pi^T \sigma(t) \, \rho \, v_{wc} \right], \cr
&  v(0, c, t) = 1, \; v(w, 0, t) = 0, \hbox{ and } \lim_{s \to \infty} \E^{w, c, t} \left[ e^{-\int_t^s \la(u) du} v(W^*(s), c(s), s) \right] = 0.} \right. \eqno(3.1)
$$
The optimal investment strategy $\pi^*$ is given in feedback form by the first-order necessary condition in $(3.1)$ subject to the constraint in $(2.3)$.  \qed}

Note that $\psi$ is homogenous of degree 0 with respect to $w$ and $c$; that is, if $k > 0$, then $\psi(kw, kc, t) = \psi(w, c, t)$.  This homogeneity implies that $\psi(w, c, t) = \psi(w/c, 1, t)$ for $c > 0$.  (Recall that for $c = 0$, we have $\psi(w, 0, t) = 0$ for $w > 0$.)  For this reason, we can reduce the number of state variables from two to one by defining $z = w/c$, and by defining the function $\phi$ on $\rp^2$ by $\phi(z, t) = \psi(z, 1, t)$.  The function $\phi$ can be interpreted as the minimum probability of lifetime ruin corresponding to some controlled wealth process; see Bayraktar and Young (2007c, Section 2).  We have the following lemma for the function $\phi$ that follows directly from Lemma 3.1.

\lem{3.2}{The minimum probability of lifetime ruin $\psi$ can be expressed as $\psi(w, c, t) = \phi(w/c, t)$, in which $\phi$ is decreasing and convex with respect to its first variable, lies in ${\cal C}^{2,1}(\rp^2)$, and is the unique classical solution of the following HJB equation on $\rp^2:$
$$
\left\{ \eqalign{&  \la(t) \, v = v_t + ((b^2(t) - a(t)) z - 1) v_z + {1 \over 2} \, b^2(t) \, z^2 \, v_{zz} \cr
& \qquad  + \min_{e^T \a = z} \left[ \a^T \mu(t) \, v_z + {1 \over 2} \, \a^T \, \Sigma(t) \, \a \, v_{zz} - b(t) \, \a^T \sigma(t) \, \rho \, (z v_{zz} + v_z) \right], \cr
&  v(0, t) = 1 \hbox{ and } \lim_{s \to \infty} \E^{z, t} \left[ e^{-\int_t^s \la(u) du} v(Z^*(s), s) \right] = 0,} \right. \eqno(3.2)
$$
in which $Z^*(t) = W^*(t)/c(t),$ with $W^*$ the optimally controlled wealth.  The optimal strategy for the HJB equation in $(3.2),$ namely $\a^*,$ is given in feedback form by the first-order necessary condition subject to $e^T \a = z$ and is related to the optimal investment strategy in $(3.1)$ by  $\pi^*(t) = c(t) \, \a^*(t)$.  \qed}

We have a mutual fund theorem that follows from Lemma 3.2, which demonstrates that the original problem in wealth and random consumption satisfies a two-fund property.  First, we define the following vectors to which we refer in the statement of the theorem.  The vectors $g$, $f$, and $h$ are defined by
$$
g(t) := {\Sigma(t)^{-1} e \over e^T \Sigma(t)^{-1} e},
$$
$$
f(t) := \Sigma(t)^{-1} \left( \mu(t) - {e^T \Sigma(t)^{-1} \mu(t) \over e^T \Sigma(t)^{-1} e} \, e \right),
$$
and
$$
h(t) := \Sigma(t)^{-1} \left( \sigma(t) \rho -  {e^T \Sigma(t)^{-1} \sigma(t) \rho \over e^T \Sigma(t)^{-1} e} \, e \right).
$$
Note that $e^T g(t) = 1$, $e^T f(t) = 0$, and $e^T h(t) = 0$ for all $t \ge 0$, so $g + k_1 f + k_2 h$ is a relative portfolio vector, as in Definition 2.1, for any deterministic functions $k_1$ and $k_2$ of time.

\th{3.3}{The optimal investment strategy $\pi^*$ to minimize the probability of lifetime ruin is to invest the dollar amount $- {\psi_w(W^*(t), c(t), t) \over \psi_{ww}(W^*(t), c(t), t)}$ in the fund defined by the relative portfolio vector $g + f,$ with the remainder of wealth in the fund defined by $g + bh$.}

\pf  As stated in Lemma 3.2, the optimal investment $\a^*$ is given by the first-order necessary condition in (3.2) subject to the constraint $e^T \a = z$, and $\a^*$ in turn gives us $\pi^*$ by $\pi^*(t) = c(t) \, \a^*(t)$.  To compute $\a^*$ in terms of $\phi$, first form the Lagrangian
$$
{\cal L} =  \a^T \mu(t) \, \phi_z + {1 \over 2} \, \a^T \, \Sigma(t) \, \a \, \phi_{zz} - b(t) \, \a^T \sigma(t) \, \rho \, (z \phi_{zz} + \phi_z)  - \ell (e^T \a - w),
$$
for some Lagrange multiplier $\ell$.  Differentiate this expression with respect to $\a$ and set the result equal to zero to obtain
$$
\a^* = - {\phi_z \over \phi_{zz}} \Sigma(t)^{-1} \mu(t) + \left( z + {\phi_z \over \phi_{zz}} \right) b(t) \, \Sigma(t)^{-1} \sigma(t) \, \rho + {\ell \over \phi_{zz}} \Sigma(t)^{-1} e. \eqno(3.3)
$$
Impose the constraint that $e^T \a^* = z$, from which it follows that
$$
\ell = {z \phi_{zz} + \phi_z \, e^T \, \Sigma(t)^{-1} \mu(t) - (z \phi_{zz} + \phi_z) \, b(t) \, e^T \, \Sigma(t)^{-1} \sigma(t) \, \rho \over e^T \Sigma(t)^{-1} e}.
$$

After substituting this expression for $\ell$ into (3.3), we get that the optimal strategy $\a^*$ in (3.2) is given in feedback form by
$$
\a^*(t) = Z^*(t) \, g(t) - {\phi_z(Z^*(t), t) \over \phi_{zz}(Z^*(t), t)} \, f(t) +  \left( Z^*(t) + {\phi_z(Z^*(t), t) \over \phi_{zz}(Z^*(t), t)} \right) b(t) \, h(t).
$$

From Lemma 3.2, we know that the optimal investment strategy $\pi^*$ is given by $\pi^*(t) = c(t) \, \a^*(t)$ and $W^*(t) = c(t) \, Z^*(t)$, in which $W^*$ is the optimally controlled wealth.  Note that $c \, \phi_z/\phi_{zz} = \psi_w/\psi_{ww}$, from which it follows that
$$
\eqalign{\pi^*(t) &= W^*(t) \left( g(t) + b(t) \, h(t) \right) - {\psi_w(W^*(t), c(t), t) \over \psi_{ww}(W^*(t), c(t), t)} \left( f(t) - b(t) \, h(t) \right) \cr
&= \left( W^*(t) + {\psi_w(W^*(t), c(t), t) \over \psi_{ww}(W^*(t), c(t), t)} \right) \left( g(t) + b(t) \, h(t) \right) - {\psi_w(W^*(t), c(t), t) \over \psi_{ww}(W^*(t), c(t), t)} \left( g(t) + f(t)\right).} \eqno(3.4)
$$
Thus, the optimal investment strategy is as stated. \qed \medskip

As a corollary to this theorem, we have the following mutual fund result in the case for which consumption is deterministic.

\cor{3.5}{When the rate of consumption is deterministic, the optimal investment strategy is to invest the dollar amount $- {\psi_w(W^*(t), c(t), t) \over \psi_{ww}(W^*(t), c(t), t)}$ in the fund defined by $g + f,$ with the remainder of wealth in the fund defined by $g$.}

\pf This result follows immediately from Theorem 3.4 by setting $b$ to be identically zero in (3.4). \qed  \medskip

Although the mutual funds in Corollary 3.5 are identical to those obtained when maximizing expected utility (that is, they lie on the same line in ${\bf R}^n$), the dollar amount invested in each fund depends on the risk preferences of the individual in each setting.

\subsect{3.2 Including the Riskless Asset}

In this section, we allow the individual to invest in a riskless asset, so we do not impose the constraint on admissible investment strategies given in (2.3), and the amount invested in the riskless asset at time $t$ is $W(t) - e^T \pi(t)$.

\lem{3.4}{For the model in this section, the minimum probability of lifetime ruin $\psi$ is decreasing and convex with respect to $w$, increasing with respect to $c$, and lies in ${\cal C}^{2,2,1}(\rp^3)$.  Additionally, $\psi$ is the unique classical solution of the following HJB equation on $\rp^3:$
$$
\left\{ \eqalign{&  \la(t) \, v = v_t + (rw - c) \, v_w + a(t) \, c \, v_c + {1 \over 2} \, b^2(t) \, c^2 \, v_{cc} \cr
& \qquad  + \min_\pi \left[ \pi^T (\mu(t) - re) \, v_w + {1 \over 2} \, \pi^T \, \Sigma(t) \, \pi \, v_{ww} + b(t) \, c \, \pi^T \sigma(t) \, \rho \, v_{wc} \right], \cr
&  v(0, c, t) = 1, \; v(w, 0, t) = 0, \hbox{ and } \lim_{s \to \infty} \E^{w, c, t} \left[ e^{-\int_t^s \la(u) du} v(W^*(s), c(s), s) \right] = 0.} \right. \eqno(3.5)
$$
The optimal investment strategy $\pi^*$ is given in feedback form by the first-order necessary condition in $(3.5)$.  \qed}

As in Section 3.1, $\psi(w, c, t) = \psi(w/c, 1, t)$ for $c > 0$, so we define $\phi$ by $\phi(z, t) = \psi(z, 1, t)$ as before.  Also, we have the following lemma that follows directly from Lemma 3.4, just as Lemma 3.2 follows from Lemma 3.1.

\lem{3.6}{The minimum probability of lifetime ruin $\psi$ can be expressed as $\psi(w, c, t) = \phi(w/c, t)$, in which $\phi$ is decreasing and convex with respect to its first variable, lies in ${\cal C}^{2,1}(\rp^2)$, and is the unique classical solution of the following HJB equation on $\rp^2:$
$$
\left\{ \eqalign{&  \la(t) \, v = v_t + ((r + b^2(t) - a(t)) z - 1) v_z + {1 \over 2} \, b^2(t) \, z^2 \, v_{zz} \cr
& \qquad  + \min_\a \left[ \a^T (\mu(t) - re) \, v_z + {1 \over 2} \, \a^T \, \Sigma(t) \, \a \, v_{zz} - b(t) \, \a^T \sigma(t) \, \rho \, (z v_{zz} + v_z) \right], \cr
&  v(0, t) = 1 \hbox{ and } \lim_{s \to \infty} \E^{z, t} \left[ e^{-\int_t^s \la(u) du} v(Z^*(s), s) \right] = 0,} \right. \eqno(3.6)
$$
in which $Z^*(t) = W^*(t)/c(t),$ with $W^*$ the optimally controlled wealth.  The optimal strategy for the HJB equation in $(3.6),$ namely $\a^*,$ is given in feedback form by the first-order necessary condition and is related to the optimal investment strategy in $(3.5)$ by $\pi^*(t) = c(t) \a^*(t)$. \qed}

We have a mutual fund theorem that follows from Lemma 3.6, which demonstrates that the original problem in wealth and random consumption satisfies a two-fund property.  First, define the two $(n+1)$-vectors $\tilde g$ and $\tilde f$ by
$$
\tilde g(t) := \left( 1 - b(t) e^T \Sigma(t)^{-1} \sigma(t) \rho, b(t) (e^T \Sigma(t)^{-1} \sigma(t) \rho)_1, \dots, (b(t) e^T \Sigma(t)^{-1} \sigma(t) \rho)_n \right)^T,
$$
and
$$
\tilde f(t) := \left( - e^T \Sigma(t)^{-1} \tilde \mu(t), (\Sigma(t)^{-1} \tilde \mu(t))_1, \dots, (\Sigma(t)^{-1} \tilde \mu(t))_n \right)^T,
$$
in which $\tilde \mu(t) := \mu(t) - re - b(t) \, \sigma(t) \, \rho$.  Note that $e^T \tilde g(t) = 1$ and $e^T \tilde f(t) = 0$ for all $t \ge 0$; here $e$ is the $(n+1)$-vector of 1's.  Thus, as in Definition 2.1, the vector $\tilde g$ defines a mutual fund that includes the riskless asset as follows:  proportion $1 - b(t) e^T \Sigma(t)^{-1} \sigma(t) \rho$ is invested in the riskless asset at time $t$, while proportion $(b(t) e^T \Sigma(t)^{-1} \sigma(t) \rho)_i$ is invested in the $i$th risky asset for $i = 1, 2, \dots, n$.  Similarly, the relative vector $\tilde g + \tilde f$ defines a mutual fund that includes the riskless asset.

The dynamics of the value of the mutual fund determined by, say $\tilde g$, is given by an expression similar to (2.6) with drift $\mu^{\tilde g}(t) = (1 - b(t) e^T \Sigma(t)^{-1} \sigma(t) \rho) \,r + \sum_{i=1}^n (b(t) e^T \Sigma(t)^{-1} \sigma(t) \rho)_i \, \mu_i(t)$ and volatility vector $\sigma^{\tilde g}(t) = \sum_{i=1}^n (b(t) e^T \Sigma(t)^{-1} \sigma(t) \rho)_i \, \sigma_i(t)$.

\th{3.7}{The optimal investment strategy to minimize the probability of lifetime ruin is to invest the dollar amount $-{\psi_w(W^*(t), c(t), t) \over \psi_{ww}(W^*(t), c(t), t)}$ in the fund defined by $\tilde g + \tilde f,$ as in the preceding discussion, with the remainder of wealth in the fund defined by $\tilde g$.}

\pf  As stated in Lemma 3.6, the optimal $\a^*$ is given by the first-order necessary condition in (3.6).  By differentiating $\a^T (\mu(t) - re) \, \phi_z + {1 \over 2} \, \a^T \, \Sigma(t) \, \a \, \phi_{zz} - b(t) \, \a^T \sigma(t) \, \rho \, (z \phi_{zz} + \phi_z)$ with respect to $\a$ and setting the result equal to zero, we obtain that the optimal strategy $\a^*$ is given by
$$
\a^*(t) = Z^*(t) \, b(t) \, \Sigma(t)^{-1} \sigma(t) \, \rho - {\phi_z(Z^*(t), t) \over \phi_{zz}(Z^*(t), t)} \Sigma(t)^{-1} (\mu(t) - re - b(t) \, \sigma(t) \, \rho),
$$
from which it follows that the optimal investment strategy $\pi^*$ is given by
$$
\eqalign{\pi^*(t) &= W^*(t) \, b(t) \, \Sigma(t)^{-1} \sigma(t) \, \rho - {\psi_w(W^*(t), c(t), t) \over \psi_{ww}(W^*(t), c(t), t)} \Sigma(t)^{-1} (\mu(t) - re - b(t) \, \sigma(t) \, \rho) \cr
&= \left( W^*(t) + {\psi_w(W^*(t), c(t), t) \over \psi_{ww}(W^*(t), c(t), t)} \right) b(t) \, \Sigma(t)^{-1} \sigma(t) \, \rho \cr 
& \qquad - {\psi_w(W^*(t), c(t), t) \over \psi_{ww}(W^*(t), c(t), t)} \left( b(t) \, \Sigma(t)^{-1} \sigma(t) \, \rho + \Sigma(t)^{-1} \tilde \mu(t) \right),} \eqno(3.7)
$$
with the remainder of wealth in the riskless asset, namely,
$$
\eqalign{W^*(t) - e^T \pi^*(t) &= W^*(t) \left(1 - b(t) \,e^T \Sigma(t)^{-1} \sigma(t) \, \rho \right) - {\psi_w(W^*(t), c(t), t) \over \psi_{ww}(W^*(t), c(t), t)} \left(- e^T \Sigma(t)^{-1} \tilde \mu(t) \right) \cr
&= \left( W^*(t) + {\psi_w(W^*(t), c(t), t) \over \psi_{ww}(W^*(t), c(t), t)} \right) \left(1 - b(t) \,e^T \Sigma(t)^{-1} \sigma(t) \, \rho \right) \cr
& \quad - {\psi_w(W^*(t), c(t), t) \over \psi_{ww}(W^*(t), c(t), t)} \left(1 - b(t) \,e^T \Sigma(t)^{-1} \sigma(t) \, \rho - e^T \Sigma(t)^{-1} \tilde \mu(t) \right).}
$$
Thus, the optimal investment strategy is as stated.  \qed  \medskip

As a corollary to this theorem, we have the following mutual fund result in the case for which consumption is deterministic.

\cor{3.8}{When the rate of consumption is deterministic, the optimal investment strategy is to invest the dollar amount $- {\psi_w(W^*(t), c(t), t) \over \psi_{ww}(W^*(t), c(t), t)} \, e^T \Sigma(t)^{-1} (\mu(t) - re)$ in the fund defined by the relative portfolio vector $\hat g(t) := {\Sigma(t)^{-1} (\mu(t) - re) \over e^T \Sigma(t)^{-1} (\mu(t) - re)},$ with the remainder of wealth in the riskless asset.}

\pf  Set $b$ identically equal to zero in (3.7) to obtain that
$$
\pi^*(t) = - {\psi_w(W^*(t), c(t), t) \over \psi_{ww}(W^*(t), c(t), t)} \Sigma(t)^{-1} (\mu(t) - re),
$$
from which the result follows.  \qed  \medskip

As for Corollary 3.5, note that $\hat g$ defines the same mutual fund that one obtains when maximizing expected utility, but the dollar amount invested in the fund depends on the risk preferences of the individual.  In this setting, the risk preferences are embodied in the probability of ruin $\psi$.

\ex{3.9} Suppose the parameters of the model are time-homogeneous, that is, constant, and suppose that the rate of consumption is constant.  In this case, the work of Young (2004) applies to give us the probability of lifetime ruin $\psi$ as follows:
$$
\psi(w) = \left(1 - {rw/c} \right)^p, \quad 0 \le w \le c/r,
$$
in which $p = {1 \over 2r} \left[ (r + \la + m) + \sqrt{(r + \la + m)^2 - 4 r \la} \right]$ and $m = {1\over 2} \, (\mu - re)^T \Sigma^{-1} (\mu -r e)$.  Also, the optimal dollar amount invested in the risky mutual fund (determined by the relative portfolio vector $\Sigma^{-1} (\mu - re) / e^T \Sigma^{-1} (\mu - re)$) is given in feedback form by
$$\pi^*(t) = {c/r - W^*(t) \over p - 1} \, e^T \Sigma^{-1} (\mu - re).$$

\sect{\bf Acknowledgement}

The research of the first author is supported in part by the National Science Foundation under grant DMS-0604491.  We thank Wang Ting for carefully reading the paper and an anonymous referee for helpful comments.

\vfill \eject

\sect{References}



\noindent \hangindent 20 pt Bayraktar, E. and V. R. Young (2007a), Correspondence between lifetime minimum wealth and utility of consumption, {\it Finance and Stochastics}, 11 (2): 213-236.

\smallskip \noindent \hangindent 20 pt Bayraktar, E. and V. R. Young (2007b), Minimizing the probability of lifetime ruin under borrowing constraints, {\it Insurance: Mathematics and Economics}, 41: 196-221.

\smallskip \noindent \hangindent 20 pt Bayraktar, E. and V. R. Young (2007c), Proving the regularity of the minimal probability of ruin via a game of stopping and control, working paper, Department of Mathematics, University of Michigan, available at Êhttp://arxiv.org/abs/0704.2244

\smallskip \noindent \hangindent 20 pt Bj\"ork, T. (2004), {\it Arbitrage Theory in Continuous Time}, second edition, Oxford University Press, Oxford.



\smallskip \noindent \hangindent 20 pt Duffie, D., W. Fleming, M. Soner, and T. Zariphopoulou (1997), Hedging in incomplete markets with HARA utility, {\it Journal of Economic Dynamics and Control}, 21: 753-782.





\smallskip \noindent \hangindent 20 pt Khanna, A. and M. Kulldorff (1999), A generalization of the mutual fund theorem, {\it Finance and Stochastics}, 3: 167-185.




\smallskip \noindent \hangindent 20 pt Merton, R. C. (1971), Optimum consumption and portfolio rules in a continuous time model, {\it Journal of Economic Theory}, 3: 373-413.

\smallskip \noindent \hangindent 20 pt Merton, R. C. (1973), An intertemporal capital asset pricing model, {\it Econometrica}, 41 (5): 867-887.

\smallskip \noindent \hangindent 20 pt Young, V. R. (2004), Optimal investment strategy to minimize the probability of lifetime ruin, {\it North American Actuarial Journal}, 8 (4): 105-126.


\bye